\documentclass[prb,preprint, showpacs,amsmath,graphicx,amssymb,superscriptaddress]{revtex4-1}

\usepackage{amsfonts}
\usepackage{amssymb}
\usepackage{amsmath}
\usepackage{graphicx}

\usepackage{subfigure}
\begin{document}

\title{Effects of Manganese Addition on the Electronic Structure of BaTiO$_3$}
\author{ J. F. Nossa}
\author{Ivan I. Naumov}
\affiliation
{Geophysical Laboratory, Carnegie Institution of Washington, Washington D.C. 
20015, USA}
\author{R. E. Cohen}
\affiliation
{Geophysical Laboratory, Carnegie Institution of Washington, Washington D.C. 
20015, USA}
\affiliation{Department of Earth Sciences, University College London, Gower 
Street, WC1E 6BT, London, United Kingdom}
\date{\today}

\begin{abstract} 
Mn is used as a dopant to improve the electromechanical 
properties of perovskite oxides. We elucidate the role of Mn defects and 
associated vacancies on the 
electronic, atomic and ferroelectric properties of BaTiO$_3$.  
Using density functional theory (DFT) and DFT+U we investigate 
the equilibrium geometry and electronic properties of Mn ion on A or B 
sites and with compensating oxygen vacancies.
We study the change in the oxidation state of Mn  
in response to local environment changes, such as the presence of oxygen
vacancies. 
\end{abstract}

 \maketitle

Transition metal dopants improve  the electromechanical 
properties of compounds like BaTiO$_3$, PbTiO$_3$  or PbZr$_x$Ti$_{1-x}$O$_3$ 
(PZT),\cite{Perez-Delfin2011, Wu2008,
Park2001, Hentati2010, Neumann1987, Arlt1988, Ren2004,Zhang2005,Zhang2006} but the origins of these effects are not well understood.
Effects of
the dopants can be divided into two categories, extrinsic and intrinsic. 
Extrinsic
effects are mainly attributed to pinning of domain walls by the 
dopant 
ions  and any associated oxygen vacancies. The intrinsic effects are not
directly connected with domain boundary movement. They manifest themselves 
through the
interaction between the doping induced local polarization $\boldsymbol{P}_{d}$ 
 and the
spontaneous polarization $\boldsymbol{P}_{s}$. It has been proposed that the 
defective dipoles
can cooperatively align along a certain crystallographic direction, producing  
a 
macroscopically observable internal bias field.\cite{Neumann1987, Arlt1988} 
One interesting 
intrinsic effect discussed in the literature is the possibility to increase 
the piezoelectric
coefficients typical for PZT and Pb(Mg$_{1/3}$Nb$_{2/3}$)$_{1−x}$Ti$_x$O$_3$ 
(PMN-PT) by a
factor of 10-40.\cite{Ren2004} This phenomenon has also been observed in aged BaTiO$_3$ samples doped with 
Mn.\cite{Zhang2005,Zhang2006}

Among the dopants, Mn is special because it is magnetic and exhibits a 
variety of 
oxidation states (Mn$^{2+}$,
Mn$^{3+}$ and  Mn$^{4+}$, and more). According to electron-spin-resonance 
(ESR) experiments, in ABO$_3$ 
perovskites, Mn 
forms neutral defects and goes in the A site as Mn$^{2+}$,   in the B 
site as 
Mn$^{4+}$  or  as  Mn$^{2+}$  with the formation  of a charge compensation 
complex 
Mn$^{2+}$-V$_{\text{O}}^{2-}$.\cite{Serway1977,Kamiya1992,Tkach2005,Tkach2006,Laguta2007}
In terms of Kr\"{o}ger-Vink  notation these 
defects can be 
denoted as   Mn$_{\text{A}}^{\times}$,    Mn$_{\text{B}}^{\times}$ and  
(Mn$_{\text{B}}^{\prime \prime}$--V$_{\text{O}}^{\bullet \bullet}$)$^ 
{\times}$,
respectively, where  the symbol  $\times$ stands for the neutral case,  
$\prime $ for the 
net negative charge,  and $\bullet$ for  net positive charge relative to a 
defect-free host.
When doped with Mn, the incipient ferroelectrics SrTiO$_{3}$ and KTaO$_{3}$ 
exhibit 
simultaneous spin and dipole glass behaviors with large non-linear 
magnetoelectric coupling.\cite{Kleemann2008,Shvartsman2008,Kleemann2009,Tkach2005}
 Such multiglass 
systems have 
attracted considerable attention because they can be viewed as a new class of 
multiferroics.

In contrast to SrTiO$_{3}$ and KTaO$_{3}$, ferroelectric  BaTiO$_3$ doped with a 3$d$ transition 
metal becomes multiferroic. This was shown theoretically within the framework of the 
local spin-density approximation (LSDA) and experimentally by magnetization measurements.\cite{Nakayama2001,Lee2003}
The ferroelectric  and magnetic  ordering in such
3$d$-metal-doped BaTiO$_3$ multiferroics  actually compete with each other, as 
demonstrated  in 
Ref.~\onlinecite{Shuai2011} by studying BaTiO$_3$:Mn thin films. It was found that only 
films with a 
large amount of oxygen vacancies exhibit room-temperature ferromagnetic 
behaviour, and which was attributed to the formation of bound magnetic 
polarons.\cite{Coey2005}

We briefly review some previous work on first-principles theory of 
defects in oxide ferroelectrics: Yao and Fu studied the formation energies of 
different   
types of  vacancies in PbTiO$_3$ as a function of (i) the  Fermi level and (ii)  
chemical 
potentials of the atomic reservoirs.\cite{Yao2011} Cockayne and Burton 
computed  the 
dipole moment of the V$_{\text{O}}^{\bullet \bullet}$--V$_{\text{B}}^{\prime 
\prime}$    
divacancy  and found  that this moment can be twice as large as the dipole 
moment per cell 
of the bulk PbTiO$_3$; such defects, therefore, can be considered as  an  important 
source of 
local polarization and local fields.\cite{Cockayne2004} He and Vanderbilt 
studied  the  
pinning of domain walls  via vacancies and revealed that the pinning effect 
can be strong.\cite{He2003} Me\u stri\'c  {\it et al.}. investigated  
Fe$^{3+}$ centers in  
PbTiO$_3$ and  found that such centers tend to replace Ti$^{4+}$ as acceptors and 
form charged 
defective associates (Fe$_{\text{A}}^{\prime}$--V$_{\text{O}}^{\bullet 
\bullet}$)$^{\bullet}$.\cite{Mestric2005} The orientation of the 
(Fe$_{\text{Ti}}^{\prime}$--V$_{\text{O}}^{\bullet \bullet}$)$^ {\bullet}$ 
defective dipole was 
found to be along the $c$-axis parallel to  the spontaneous 
polarization. DFT calculations  
for Cu 
impurities in PbTiO$_3$  found  them to behave similarly to their  counterpart 
Fe acceptor centers.\cite{Eichel2008}   When isolated, their most stable charge state is 
Cu$_{\text{Ti}}^{\prime\prime}$, 
which leads to two holes in the valence band. Like Fe, the Cu centers  have 
strong chemical 
driving force for association with oxygen vacancies to form 
(Cu$_{\text{Ti}}^{\prime\prime}$--V$_{\text{O}}^{\bullet \bullet}$)$^ 
{\times}$ defective 
dipoles whose preferable orientation is again along the [001] 
axis.\cite{Eichel2008} Zhang 
{\it et al.} explored the effects of acceptor substitutes to the Ti site in  
PbTiO$_3$.\cite{Zhang2008}
They screened   group IIIB  and VB  elements and found that 
these elements  favor immobile acceptor-oxygen-vacancy-acceptor defect 
clusters. Moreover,  
they found that groups IIIB and VB dopants take two distinct 
defect-cluster 
structures along the $z$-direction and in the $xy$-plane, respectively. While 
the former 
enforces the domain pinning, the latter makes the domain pinning effects 
weaker.

\begin{figure}[t] 
\centering
\includegraphics[scale=0.5]{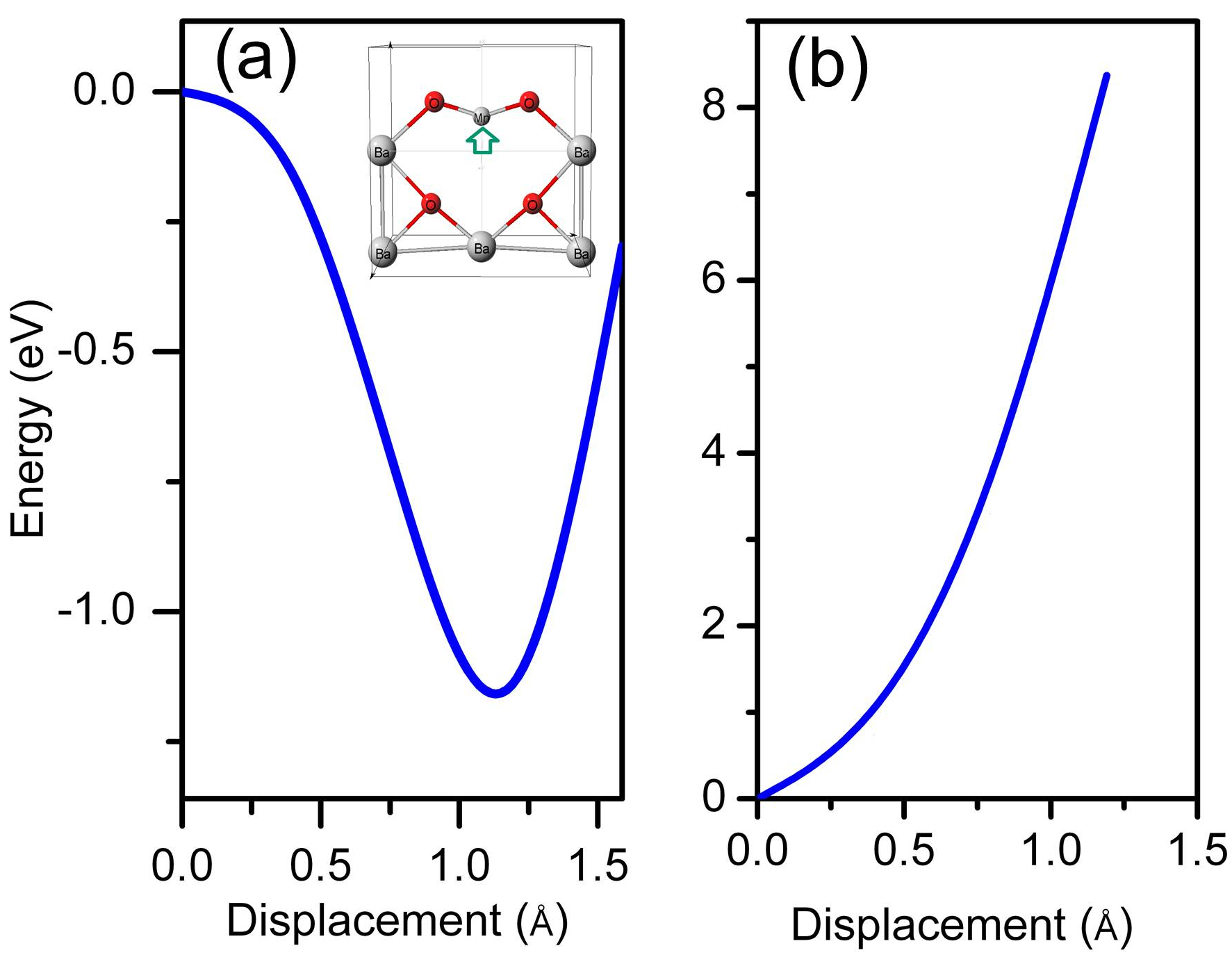}
\caption{Total energy dependence on the  Mn (a) and Ba (b) displacements along the
[001] direction in a  Ba$_{7}$Ti$_{8}$MnO$_{24}$, supercell and pure BaTiO$_3$, respectively.
The insert shows the equilibrium position of the  Mn atom in the (010) BaO plane.}
 \label{fig:fig1}
\end{figure}

Kvyatkovskii investigated  the 
geometry and electronic structure for defects produced by the Mn impurities 
both in SrTiO$_3$ and 
KTaO$_3$.\cite{Kvyatkovskii2009,Kvyatkovskii2012} He used a cluster approach where 
the crystalline environment of the defect is modeled 
by a cluster that is passivated by hydrogen atoms. Kondakova {\it et al.} in 
their local 
spin density approximation (LSDA) calculations showed that a Mn ion 
substituting for Sr in 
SrTiO$_3$ occupies the off-central position.\cite{Kondakova2009} This result 
supports the idea 
that the observed dielectric anomalies in Sr$_{1-x}$Ti$_x$O$_{3}$ are due 
to motion of dipoles associated with the off-centered 
Mn$^{2+}$
impurities.\cite{Tkach2005} Using LSDA+U and many-body perturbation theory, 
Kizian  
\emph{et al.}  
calculated magnetic interactions in doped SrTiO$_3$ between the Mn impurities of 
different kinds: 
Mn$_{\text{Sr}}^{2+}$--Mn$_{\text{Sr}}^{2+}$, 
Mn$_{\text{Sr}}^{2+}$--Mn$_{\text{Ti}}^{4+}$ 
and Mn$_{\text{Ti}}^{4+}$--Mn$_{\text{Ti}}^{4+}$.\cite{Kuzian2010}  They 
found that the 
exchange interaction between Mn$_{\text{Sr}}^{2+}$ impurities is significantly 
smaller that
those for Mn$_{\text{Ti}}^{4+}$ and 
Mn$_{\text{Ti}}^{4+}$--Mn$_{\text{Ti}}^{4+}$ pairs. The
authors concluded that the presence of Mn$_{\text{Sr}}^{2+}$ ions is necessary 
but
not sufficient to explain the multiglass behaviour in Mn-doped SrTiO$_3$;  the presence 
of  Mn$_{\text{Ti}}^{4+}$ ions is also necessary.

Our calculations have been performed using a $2 \times 2 \times 2$ supercell 
(40 atoms plus vacancies). First, we replaced a Ti atom by a Mn ion to form a 
tetragonal BaTiO$_3$:Mn structure with polarization axis along the 
$z$-direction. Then, we removed an oxygen atom to form  a 
BaTiO$_3$:Mn-V$_{\text{O}}$
complex, a BaTiO$_3$ with a Mn impurity and an O vacancy. 
To study the effect of the vacancy position, 
we also have removed an oxygen atom along the {\it x}-axis. 

We have performed computations with the \textsc{abinit} code.\cite{Xavier2009,Gonze2009} 
Norm-conserving pseudopotentials and Projector-Augmented Wave atomic data have been first 
generated with \textsc{opium}  and \textsc{atompaw} codes, and then used along with the Wu-Cohen 
exchange and 
correlation functional.\cite{Wu2006} A $4\times4\times 4$ Monkhorst-Pack 
k-point grid has 
been exploited for the calculation of ground state properties. For the 
plane-wave expansion 
of the valence and conduction band wave-functions, a cutoff energy of 544 eV 
was chosen. 
To compare the effect of the large Coulombic repultion between localized electrons,
 we have used LSDA and DFT+U  approximations. In the latter, the Mn 
on-site 
Coulomb parameters, $U$=8.0 eV and $J=U/10$, were chosen. DFT+U  calculations 
give somewhat larger ($\sim$5 \%) distance between the Mn ion  and  nearest O atom 
in the structures 
containing Mn-V complexes. 

\begin{figure}[tp] \centering
\includegraphics[scale=0.11]{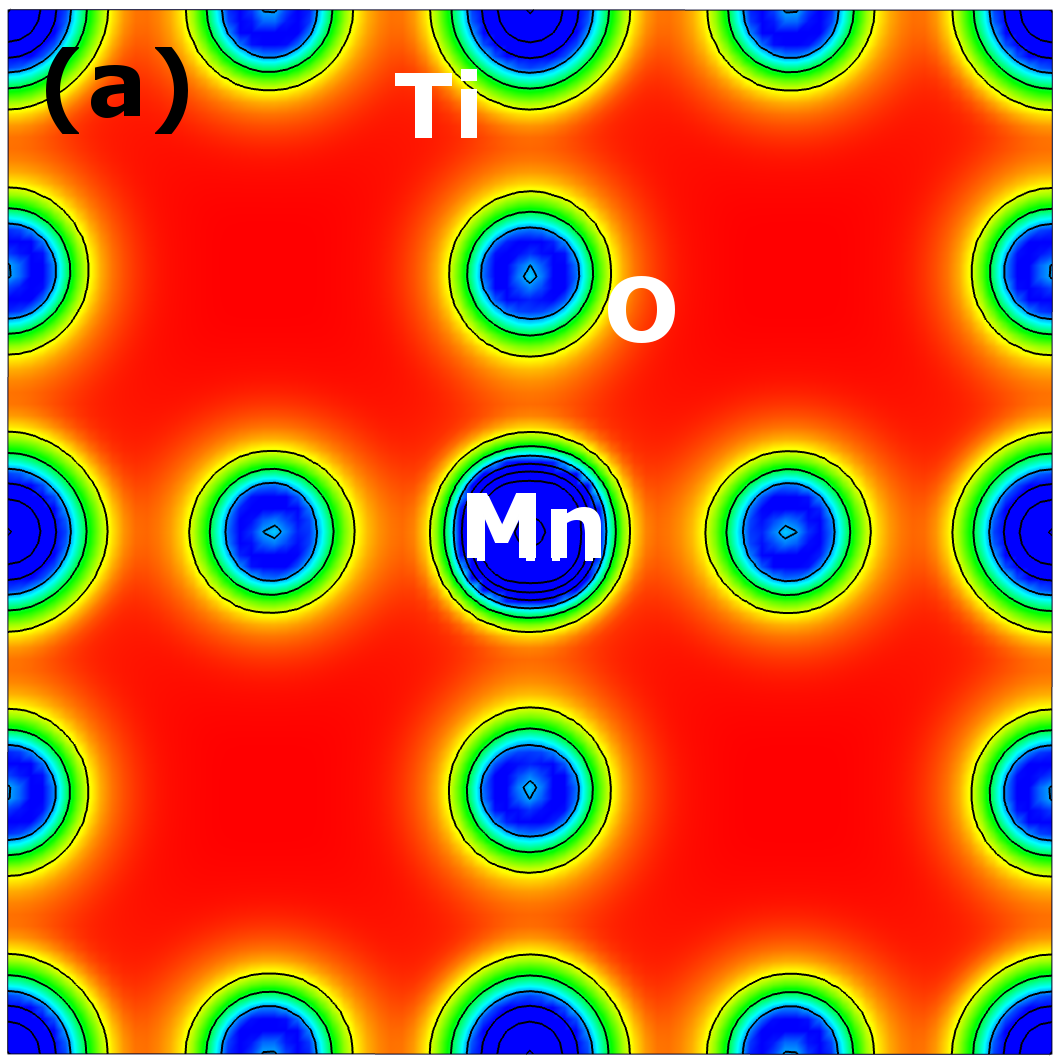}
\includegraphics[scale=0.1]{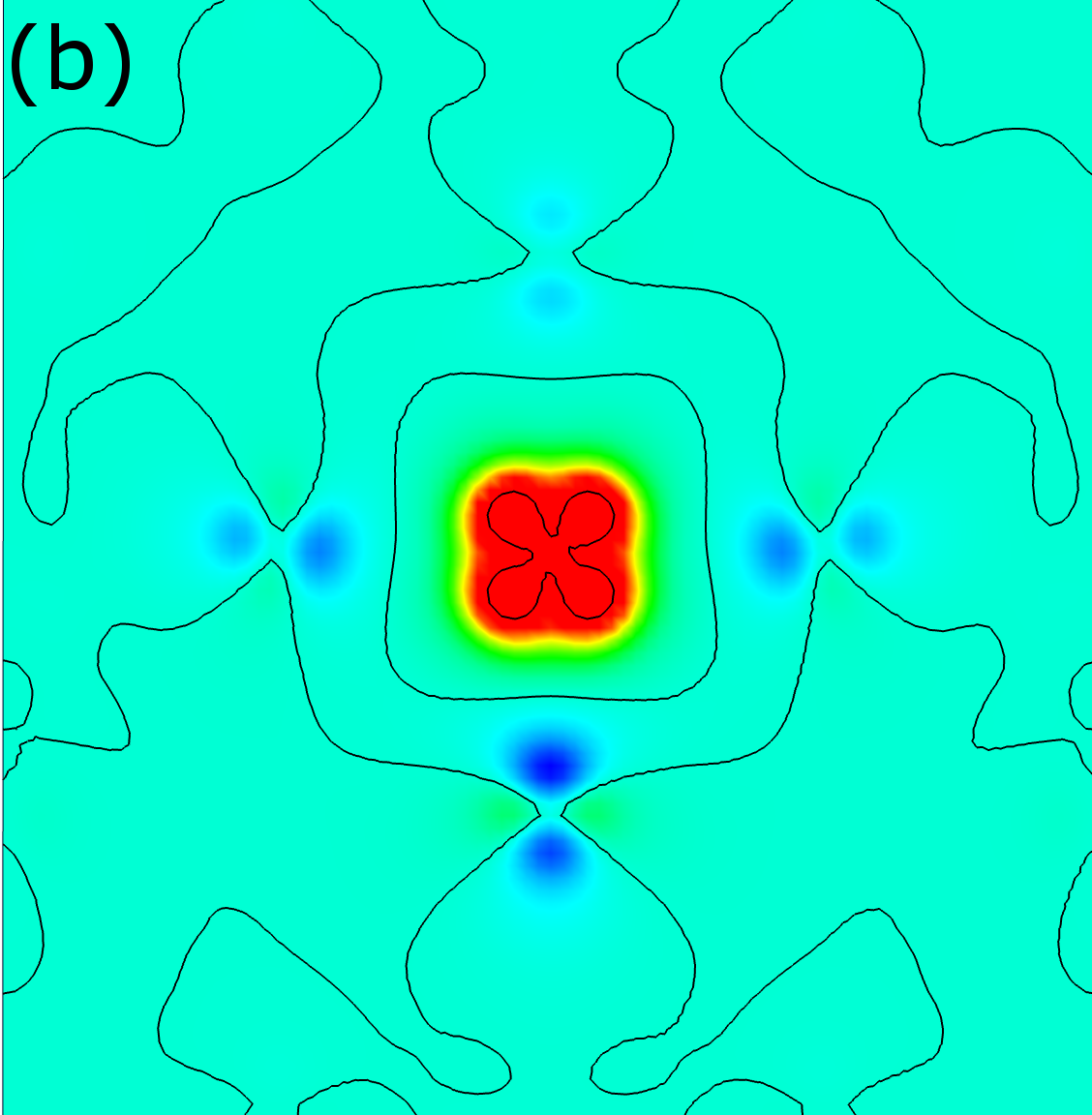}
\caption{(color online) The  valence charge (a) and spin (b)  densities around 
the  Mn$_{\text{Ti}}$ impurity in tetragonal BaTiO$_3$ in the plane parallel 
to the  spontaneous polarization.}
 \label{fig:fig2}
\end{figure}

 \begin{table}
  \caption{ Interatomic distances, $d$,  in pure tetragonal  BaTiO$_{3}$ and   with  
  Mn$_{\text{Ti}}$-V$_{\text{O}}^z$  and
 Mn$_{\text{Ti}}$-V$_{\text{O}}^x$  defect complexes.  Note that due to the breaking of cubic 
 symmetry there  are  two kinds of
   apical  (O$_{1}$ and O$_{1}^\prime$) oxygen atoms  with respect   to the $c$-axis.  In 
   the presence of  the V$_{\text{O}}^x$  vacancy  the  tetragonal
   symmetry is also broken, so that two  kinds of  equatorial (two O$_{2}$  plus one  
   O$_{2}^\prime$) oxygen atoms can be distinguished. Labels are shown in Fig. \ref{fig:labels}.
   Results from LDA+U calculations are shown in square brackets.}
\begin{ruledtabular}
\begin{tabular}{c|c|c|c}
$d$ (\AA) &pure BaTiO$_{3}$&with Mn-V$_{\text{O}}^z$  &with Mn-V$_{\text{O}}^x$ \\
 \tableline
Mn(Ti)-O$_{1}$     &2.234 & 2.013[2.069]& 2.209[2.070]\\
Mn(Ti)-O$_{1}^\prime$&1.839 &-            & 1.925[1.923]\\
  Mn(Ti)-O$_{2}$     &1.989 & 2.049[2.108]& 2.009[1.953]\\
Mn(Ti)-O$_{2}^\prime$&  -   &  -          & 1.894[2.082]\\
   \end{tabular}
   \end{ruledtabular}
   \label{tab:table1}
  \end{table}

\begin{figure}[tpbh] \centering
\includegraphics[scale=0.3]{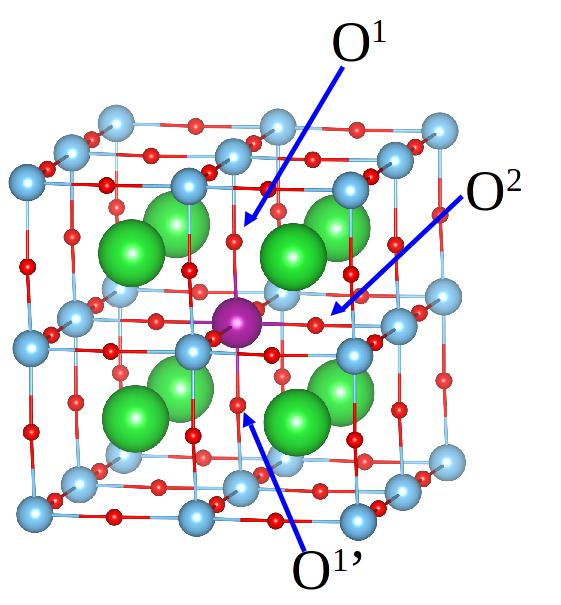}
\caption{(color online) Mn-doped BaTiO$_3$ structure. Arrows 
show two apical and one equatorial oxygen atoms, which are labeled as O$_{1}$, O$_{1}^\prime$ and O$_{2}$, respectevely. 
Purple, red, blue and green balls correspond to Mn, Oxygen, Ti, and Barium atoms, respectively} 
 \label{fig:labels}
\end{figure}

We first consider the case for a 
Mn impurity occupying the Ba site. 
In Fig.~\ref{fig:fig1}(a)  we present the ground state energy as a 
function of the value of Mn 
shift along the [001] direction. All the other atoms are fixed 
in ideal 
perovskite positions. The curve has a minimum corresponding to 1.12 \AA, which 
is twice as
large as that found in SrTiO$_3$.\cite{Kondakova2009} The same displacement of Ba in 
pure BaTiO$_3$, 
Fig.~\ref{fig:fig1}(b),  leads to a steep rise in energy. Thus, the Mn 
impurity tends
to be off-centered, like in SrTiO$_3$.\cite{Kvyatkovskii2009} Using DFT+U, we find that the Mn ion in the A site in the cubic structure has a
magnetic moment of 5.1  $\mu_B$. This indicates  that Mn$_{\text{Ba}}$
has a spin value of 5/2. This is in agreement with the 
experimental 
fact that in the incipient (cubic) ferroelectrics, like SrTiO$_3$, 
Mn$_{\text{A}}$  has a valency of 2+ and spin moment S=5/2.\cite{Laguta2007}  Decreasing  $U$ to $\sim$ 4.0 eV reduces the computed moment to 4.5  $\mu_B$. 
It is interesting that  such a relatively high   spin  state of the  
Mn$_{\text{Ba}}$ persists  in 
going from the cubic to tetragonal phase,   despite the fact that the atomic 
volume of the cubic 
phase is significantly  lower (by 4 \%) than  that in the tetragonal phase. 
This situation is in 
contrast to  what  occurs in MnO,  where applied pressure  induces
magnetic collapse,  as discussed in Ref.~\onlinecite{Cohen1997, Kunes2008}.
 
Mn in the B-site reduces $c/a$ from 1.026 to 1.020. The Mn valence charge density is
similar to the Ti, and the spin density is concentrated  mostly on the Mn ion.(Fig.~\ref{fig:fig2}). The Mn  magnetic moment is about 3.38
$\mu_B$ with DFT+U, suggesting that the  Mn$_{\text{Ti}}$  ion  has   a  
high spin value  of 3/2 and  an oxidation state   of 4+ in this case. 

We also consider two types  of Mn$_{\text{Ti}}$-V$_\text{O}$  complexes containing  two different 
kinds  of nearest oxygen vacancies--V$_{\text{O}}^z$ and  V$_\text{O}^x$ 
replacing  the 
apical  O$^{1}$ and equatorial O$^{2}$ oxygen atoms, respectively. The 
interatomic 
distances, $d$, in pure tetragonal BaTiO$_3$ and in the presence of such 
defects are listed in Table~\ref{tab:table1}. Fig. \ref{fig:labels} shows the
two apical and one equatorial oxygen atoms around Mn ion, which are labeled as 
O$_{1}$, O$_{1}^\prime$ and O$_{2}$, respectively.
We found that when the complex 
Mn$_{\text{Ti}}$--V$_{\text{O}}^z$ is introduced  
the tetragonal $c/a$ is 1.020, the same as with Mn without a vacancy.

\begin{figure*}[tpbh] \centering
\includegraphics[scale=0.45]{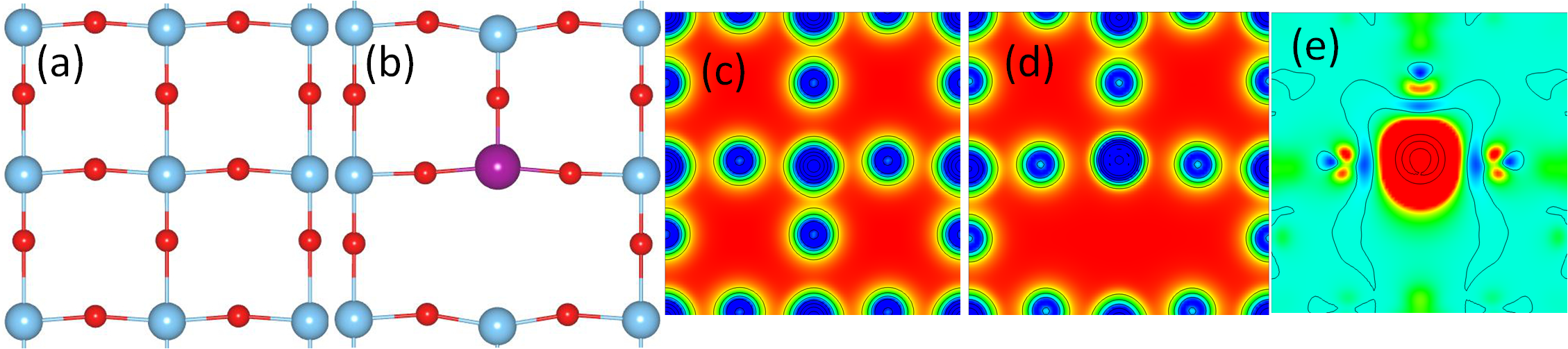}
\caption{(a) (color online) Undoped  reference  structure of  BaTiO$_3$. (b)  Mn$_{\text{Ti}}$-V$_\text{O}$ 
defective  associate oriented along the
$c$-axis. Purple, red and blue balls correspond to Mn, Oxygen and Ti atoms, respectively. Note that  Mn as  a cation  is 
displaced away  from the oxygen vacancy due to electrostatic repulsive interaction.   (c, 
d)  Valence  charge density for the structures (a)
and (b). (e)  Spin density for structure in (b),  $\rho(r)\uparrow-\rho(r)\downarrow$.}
 \label{fig:fig3}
\end{figure*}

When the defect is oriented 
along the $c$-axis or polarization direction, the displacement of the Mn ion 
is reduced from that of the Ti atoms in pure BaTiO$_3$, as seen from Table~\ref{tab:table1}
and  Figs.~\ref{fig:fig3}(a) and (b).  Valence charge densities 
for pure BaTiO$_3$ and BaTiO$_3$:Mn-V$_{{\text O}}$ are shown in Figs.~\ref{fig:fig3}(c) 
and (d), respectively. There is 
practically no charge 
density at the oxygen vacancy. The spin density (Fig.~\ref{fig:fig3}(e)) is centered around the Mn defect as expected, but overlaps with the neighboring O ions.  

Our calculations show that in  the Mn$_{\text{Ti}}$--V$_{\text{O}}^z$ complex, 
the Mn magnetic moment is 4.93 $\mu_B$ indicating that the ion  has a charge 
of 2+ with 
S=5/2. In the case of 
the Mn$_{\text{Ti}}$--V$_\text{O}^x$  defects the Mn moment is about 4.9 
$\mu_B$ consistent with a charge of 2+ and  S=5/2.  In   both cases, the Mn 
ions maintain their valence 
states, their 3$d$ shells are half-filled. 

\begin{figure}[tpbh] 
\includegraphics[scale=0.125]{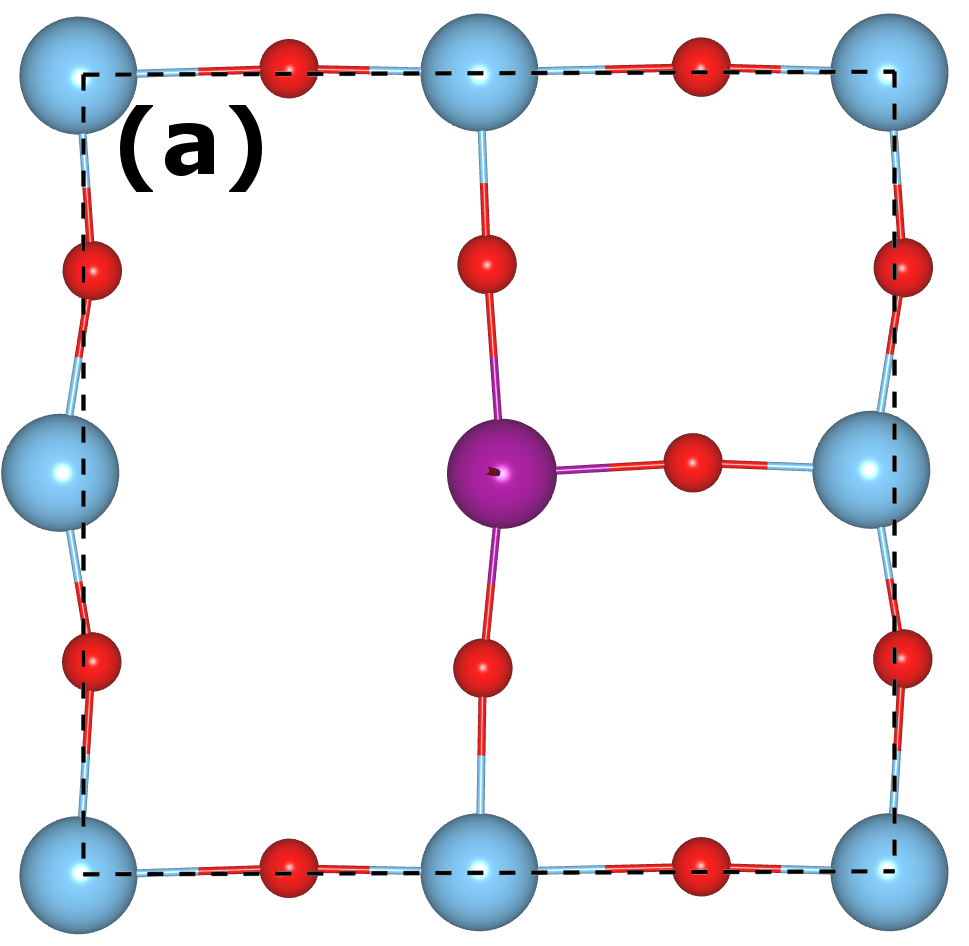}
\includegraphics[scale=0.1]{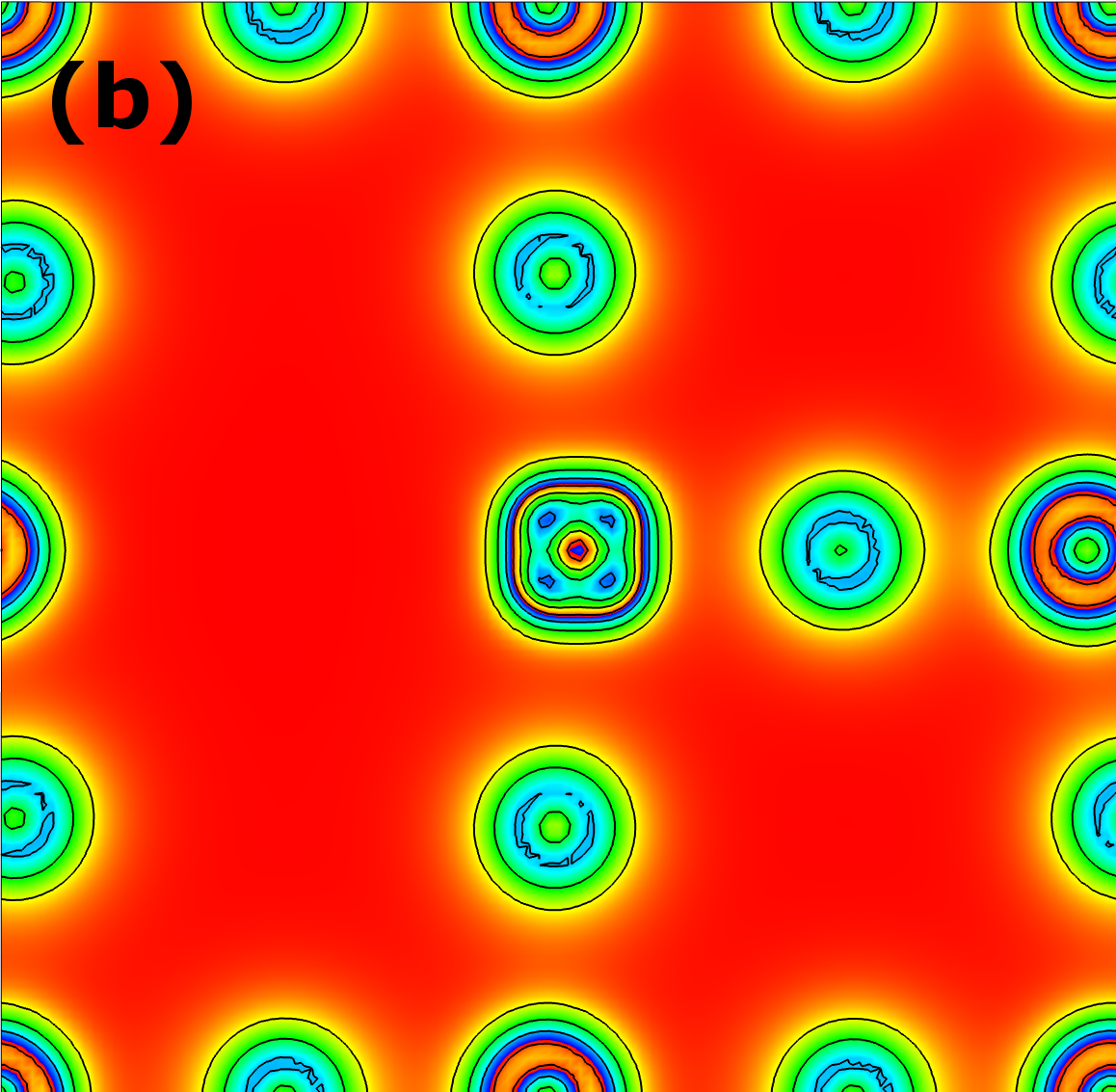}
\caption{(a) (color online) Mn$_{\text{Ti}}$--V$_\text{O}^x$ defect  with a 
vacancy occupying the nearest equatorial  oxygen position. (b) The
corresponding valence  charge density.}
\label{fig:fig4}
\end{figure}

When the vacancy occupies the nearest equatorial position relative to Mn, both 
the Mn ant Ti 
atoms move away from the vacancy, see Table~\ref{tab:table1} and  
Fig.~\ref{fig:fig4}(a). 
Again, practically no charge density is detected on the vacancy position, see 
Fig.~\ref{fig:fig4}(b). Since this defect pair is oriented 
perpendicular to the 
spontaneous polarization, it is expected to affect the ferroelectricity to a 
lesser extent than when 
oriented along the $c$-axis.

Depending on the environment,  Mn ions 
in BaTiO$_{3}$ can exist
in the oxidation  states  Mn$^{2+}$ and Mn$^{4+}$ in agreement with the direct 
ESR 
experiments.\cite{Kamiya1992} At the same time their magnetic
moments can be either 3 to 5 $\mu_B$. When a Mn impurity goes 
in the A site, the ratio
$c/a$ increases. When an Mn goes in the B site, c/a and the ferroelectric distortion decrease. 
The Mn oxygen vacancy complex with the oxygen vacancy on 
the $c$-axis is favored in energy. In particular, the V$_\text{O}$  in the 
nearest
apical position is favored by 0.10 eV  as compared to the equatorial one. 
Therefore, the probability of 
finding an oxygen vacancy around the Mn center is highest along the the 
$c$-axis. 

This work was supported by the Center the Office of Naval Research (ONR) grants grants N00014-12-1-1038 and N00014-14-1-0561
and the European Research Council Advanced Grant ToMCaT. Computations were performed on the DOD 
High Performance Computing Centers.

\bibliography{BTO}

\end{document}